\documentclass[12pt]{article}

\usepackage{mathptmx}
\usepackage[a4paper, margin=2.5cm]{geometry}

\usepackage{doi}
\usepackage{hyperref}
\hypersetup{colorlinks,citecolor=blue,urlcolor=blue}
\usepackage[neveradjust]{paralist}
\usepackage[parfill]{parskip} 

\usepackage[
    backend=biber,
    style=alphabetic,
    sortlocale=en_US,
    minnames=1, 
    maxnames=10,
    doi=true,
    eprint=false,
    autocite=inline
]{biblatex}
\AtEveryBibitem{%
  \clearfield{note}%
  \clearfield{language}%
  \clearfield{issn}%
  \clearfield{urldate}%
  \clearfield{urlyear}%
  \clearfield{urlmonth}%
  \iffieldundef{doi}{}{\clearfield{url}}%
}

\DeclareSourcemap{
  \maps[datatype=bibtex]{
    \map{
      \step[notfield=author, final]
      \step[fieldsource=note, final]
      \step[fieldset=label, origfieldval, final]
    }
  }
}

\usepackage[normalem]{ulem}

\usepackage{graphicx}
\usepackage{textcomp}
\newcommand{\tsup}[1]{\textsuperscript{#1}}

\newcommand{\tdeg}{\textdegree}
\newcommand{\tdot}{\textperiodcentered}
\newcommand{\tarr}{\textrightarrow}

\title{Czochralski growth of tin crystals as a multi-physical model experiment}
\author{Kaspars Dadzis\\
\itshape\small Leibniz-Institut für Kristallzüchtung (IKZ), Max-Born-Str. 2, 12489 Berlin, Germany}
\date{}

\begin{document}

\maketitle


\section*{Abstract}

A new setup for Czochralski growth of model materials in air atmosphere has been developed. It includes various in-situ 
measurements to access the basic physical phenomena on a macroscopic level: heat transfer, electromagnetism, melt and gas 
flows, crystal stresses. A reference experiment with tin is performed and analyzed using simple analytical 
estimates as well as 2D numerical simulations with open source models. This study aims to improve the basic physical 
understanding of the Czochralski growth process and to provide a useful tool for education and research, both for non-specialists 
and scientists.


\section{Introduction}\label{sec:intro}

In 1916 Jan Czochralski performed the famous experiment when he (presumably) dipped a pen into a crucible with molten tin 
and observed a thin thread of solidified metal hanging at the pen when pulled out \autocite{czochralski_neues_1918}. This 
relatively simple principle, initially applied to measure the maximum solidification rates of metals, has evolved to one of the main 
methods for the production of crystalline materials from the molten phase today. The modern Czochralski (CZ) growth process 
has gone through many development steps \autocite{uecker_historical_2014}, and a modern growth furnace for the production 
of silicon crystals with diameters up to 450 mm in the industry \autocite{anttila_czochralski_2020} can be hardly compared to 
Czochralski’s setup in 1916. Nevertheless, a \textit{set of physical phenomena} can be selected which is present in every growth 
experiment based on the CZ method. The aim of the present study is to define this set and to evaluate our qualitative and 
quantitative understanding of its components. In Sec.~\ref{sec:exp}, a modern version of Czochralski’s historic 
experiment is developed, still retaining the character of a simple \textit{desktop} demonstration, but adding (more) 
automatization and in-situ observation (measurements). After an application of the new setup to grow a tin crystal in 
Sec.~\ref{sec:growth}, simple analytical and numerical models are built in Sec.~\ref{sec:analyt} and \ref{sec:num}, and 
discussed in Sec.~\ref{sec:disc}.

The present study follows the concept of \textit{model experiments} for the \textit{validation} of physical models as already 
discussed for crystal growth processes in \autocite{dadzis_directional_2016}. This concept can be summarized shortly as follows: 
make the \uline{essential} physics accessible for in-situ observation (for crystal growth: make the growth furnace 
\textit{transparent}). Physical models form the foundation of numerical simulation, which is a very useful tool for process 
optimization allowing to reduce the still significant need for trial-and-error in practice \autocite{derby_modeling_2010}. Here, 
the first iteration of model experiments for the CZ process is demonstrated, where the resulting simple, low-cost setup is useful 
also for demonstration and teaching purposes. Nevertheless, the setup serves as a convenient platform for testing of in-situ 
measurement equipment. The second iteration has been already realized in terms of a small vacuum furnace 
\autocite{enders-seidlitz_model_2022}. This work is a part of the NEMOCRYS project devoted to the development of the next 
generation of multi-physical models for crystal growth \autocite{noauthor_next_nodate}.


\section{Experimental setup}\label{sec:exp}

\begin{figure}[h]
\centering
\includegraphics[width=0.95\linewidth]{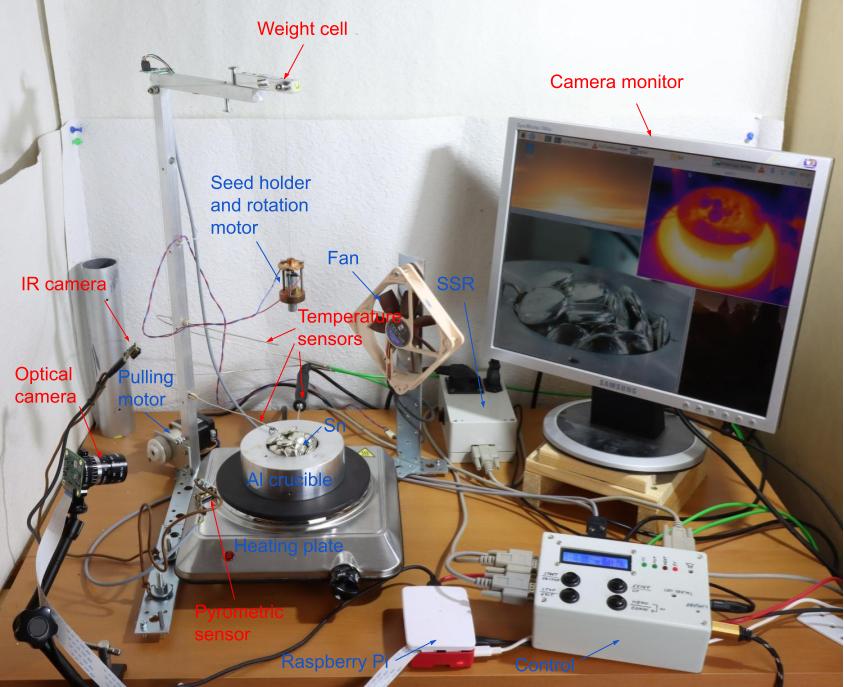}
\caption{Setup for CZ model experiments: photograph incl. control unit and sensors.}
\label{fig:setup}
\end{figure}

A simple experimental setup for CZ growth from the melt at temperatures up to 350~\tdeg C in an air atmosphere has been 
developed in this study. A photograph with the components of the setup is shown in Fig.~\ref{fig:setup} The crucible with outer 
dimensions of D120x40~mm\tsup{2} and inner dimensions of D60x25~mm\tsup{2} is made of aluminum. It is placed on a 
hotplate with a top of cast iron and 180~mm diameter. A stand consisting of two rectangular aluminum bars of 500~mm and 
200~mm length allows to guide a thin metallic thread from a seed holder with a rotation motor to a pulling motor at the bottom. 
The distance between the crucible and the top bar of 400~mm and the height of the seed holder of 60~mm leads to a maximum 
crystal length of approximately 330~mm. The base of the stand is made of flat connectors, which are also used to attach a fan 
and several sensors. The hotplate and crucible can be optionally covered by a fused silica cup (with a 80~mm or 30~mm hole on 
the top) with a diameter of 170~mm and height of 55~mm. The whole setup can be optionally covered by a 
500x500x600~mm\tsup{3} box made of acrylic glass.

Vertical motion of the seed holder is realized with a stepper motor and a 1:5.18 planet gear box. A thread wheel with a diameter 
of 30~mm enables vertical velocities up to about 10 mm/s. However, continuous motion is limited by the minimum motor step of 
1.8\tdeg, which translates to a vertical length of 0.1~mm. Seed rotation by a small DC motor including a miniature gear box is 
possible in the range 0.5\ldots 20 rpm. A fan with a diameter of 120~mm and variable rotation speed in the range 300\ldots 
1200 rpm is applied for the cooling of the crystal. 

The measurement system consists of the following components:\vspace{-\parskip}
\begin{compactitem}
\item Type K thermocouples immersed into the melt and approximately 10~cm above the crucible;
\item PT100 temperature sensors inserted in a crucible hole and apart from the setup measuring ambient air temperature;
\item Pyrometric sensor aimed on a sticker with known emissivity (0.95) on the crucible side;
\item Load cell measuring the weight on the pulling wire with a maximum load of 1~kg and resolution of 0.1~g;
\item Current sensor to determine the power consumed by the hotplate;
\item Longwave (8--14~\textmu m) infrared camera to observe the temperature distribution on the surfaces of hotplate, 
crucible, and crystal;
\item Visible light camera to observe the growth processes and the meniscus in particular.
\end{compactitem}

The hotplate and its temperature stability has a crucial role for the growth process. While there are many lab hotplates available 
with already integrated temperature control, it turned out that they can produce temperature oscillations up to 10~K (see 
App.~\ref{app:online}). Therefore, a common household hotplate with a resistive heater (power rating of 1500~W) was 
combined with a PID controller running on an Arduino microcontroller. An on/off mode with a period of 20~s, and on-time in the 
range 2\ldots 10 s was implemented for the relay. Stability of crucible temperature of 1~K was achieved up to about 
350~\tdeg C. Higher stability would probably require control of supplied power using a thyristor.

All hardware components are controlled by an Arduino microcontroller and Raspberry Pi microcomputer, see 
App.~\ref{app:online} for further details. The setup described in this section is being published with open hardware and open 
source software. The cost of the experimental setup does not exceed 1,000~\texteuro, where approximately a half was needed 
for sensors. Therefore, this simple CZ setup could be used also for various demonstration and educational purposes. 


\section{Growth experiment}\label{sec:growth}

\begin{figure}[h]
\centering
\includegraphics[width=0.85\linewidth]{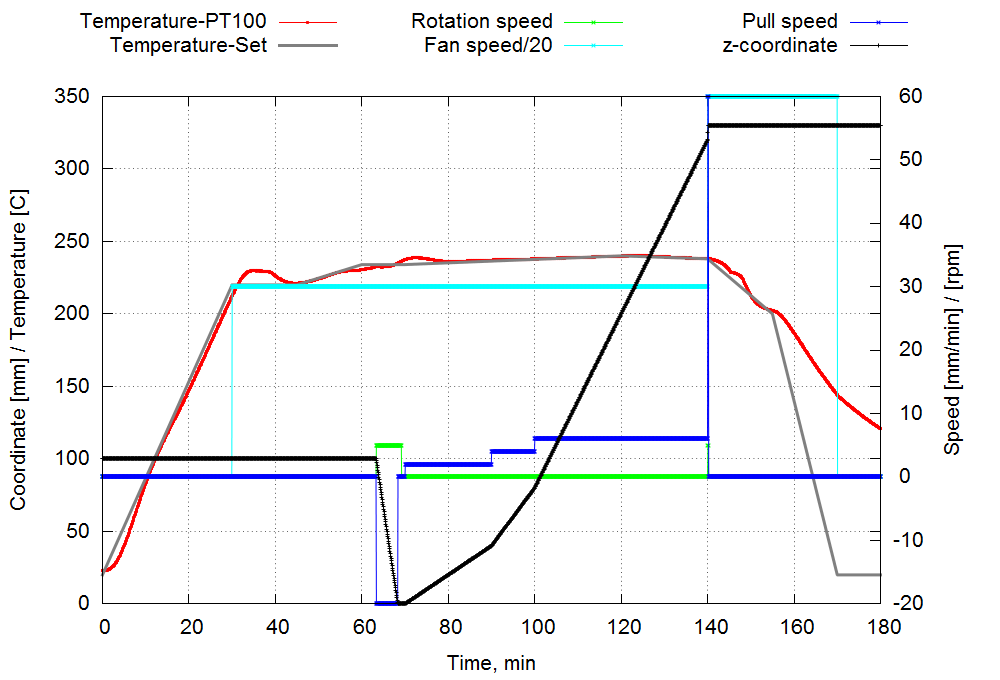}
\includegraphics[width=0.1\linewidth]{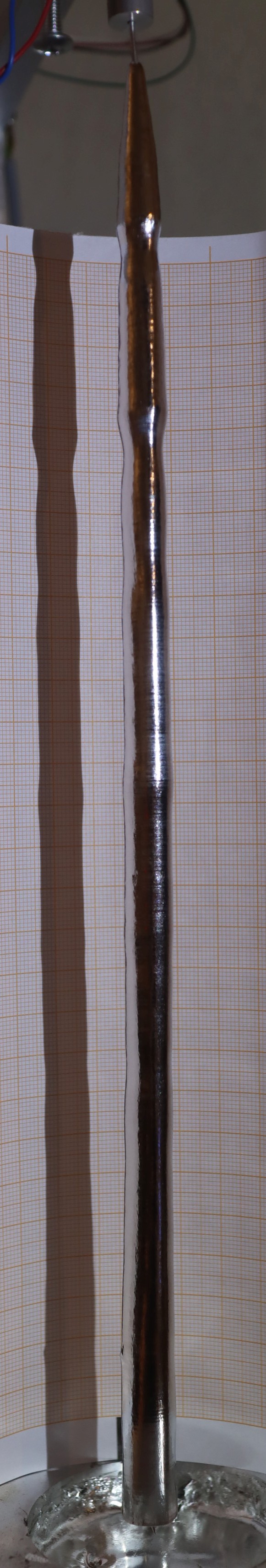}
\caption{Plot of the process data (left) and photo of the grown Sn crystal (diameter 10\ldots 13~mm) with a shadow on a scale 
paper (right).}
\label{fig:exp}
\end{figure}

In this section, a CZ growth experiment with tin (Sn) is described. Tin pellets of 99.9\% purity were used as raw material for the 
growth experiment. A seed was cut out of a thin Sn crystal with 1.3 mm diameter from a previous pulling experiment with a high 
pull speed. A filled  crucible as shown in Fig.~\ref{fig:setup} was heated following temperature ramps and arriving at 
234~\tdeg C with molten tin after 60 min (see Fig.~\ref{fig:exp}). After mechanically removing the oxide layer from the melt 
surface, the seed was lowered from its initial position 100~mm above the crucible and dipped into the melt. The growth process 
started with a pull speed of 2 mm/min for the first 20~min, followed by 4~mm/min for 10~min and 6~mm/min for the last 
40~min. Crucible temperature was increased from 234~\tdeg C at growth start to 240~\tdeg C within 50~min. The cooling fan 
was running at 600~rpm, from the end of melting. The entire process was running automatically according to a predefined recipe, 
see App.~\ref{app:online} for further details. This recipe allowed to obtain a crystal with an approximately constant diameter of 
10\ldots 13~mm as shown in Fig.~\ref{fig:exp}. With constant pull rate and crucible temperature, crystal diameter increased 
over length, it was significantly smaller without the cooling fan. The cross-section of the crystal was lens-shaped, which is  
probably caused by crystallographic orientation of an (at least) partly single crystal and by thermal melt inhomogeneities. Crystal 
rotation generally enhanced round sections in other experiments. A more detailed study including crystal characterization is 
planned later.

\begin{figure}[h]
\centering
\includegraphics[width=0.95\linewidth]{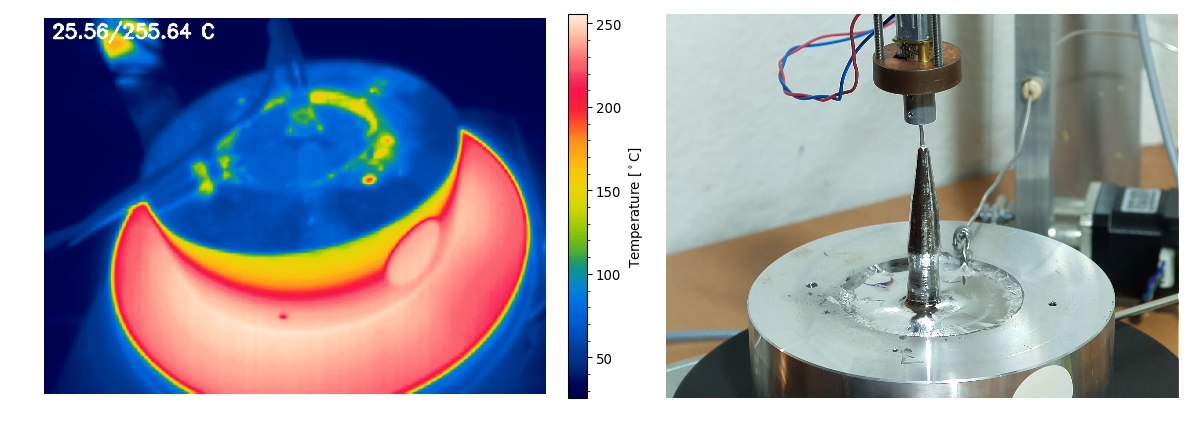}
\includegraphics[width=0.47\linewidth]{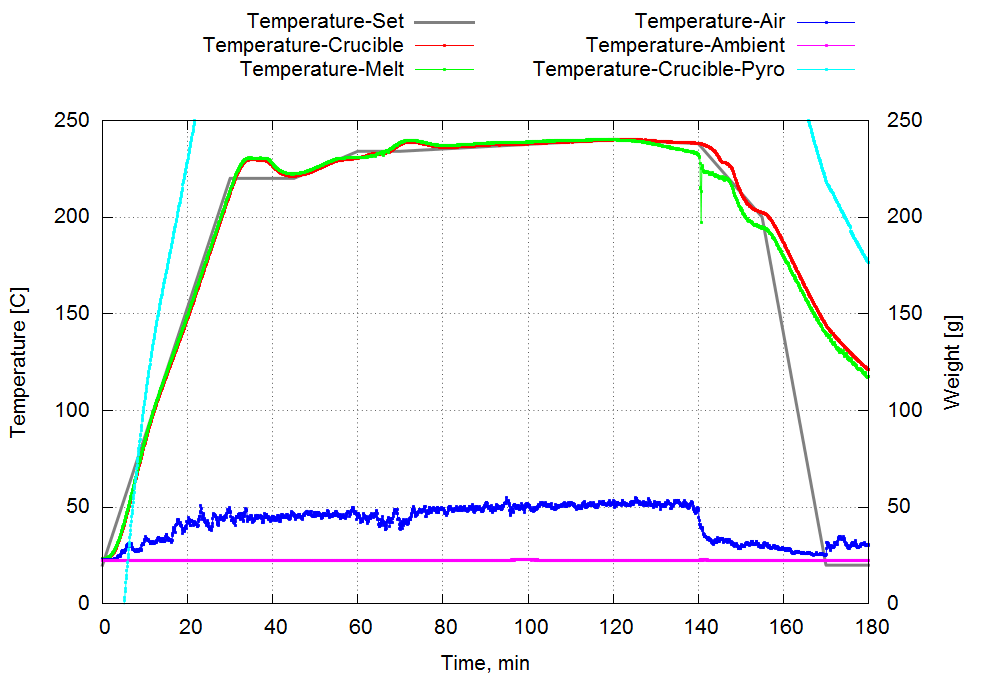}
\includegraphics[width=0.47\linewidth]{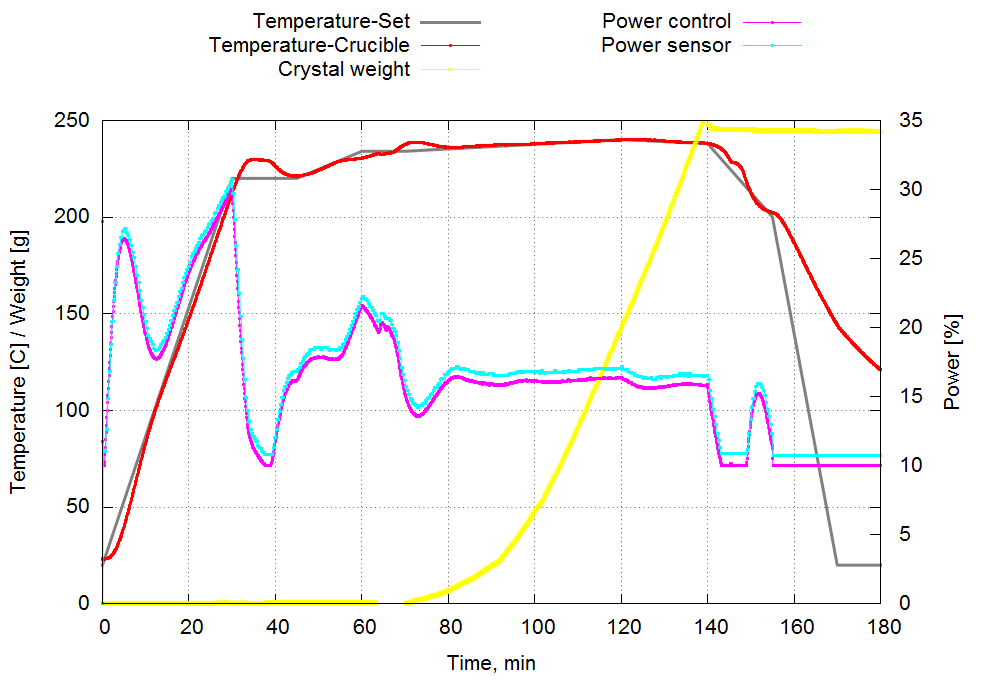}
\caption{Further in-situ measurements: IR \& optical image (top) and various sensor data (bottom).}
\label{fig:meas}
\end{figure}

Other sensor measurements in addition to crucible temperature are summarized in Fig.~\ref{fig:meas}. Melt temperature was 
measured about 1~K higher than crucible temperature, a larger deviation occurs after 120~min when the thermocouple 
becomes only partly covered by the melt. Air above the crucible reaches 50~\tdeg C, with flow-induced oscillations in an interval 
of about 5~\tdeg C, while the ambient temperature stays at 22.5\textpm 0.3~\tdeg C. The pyrometric measurement of crucible 
temperature using an emissivity sticker was not successful due to sensor malfunction not applying the correct emissivity. The 
infrared camera image shows a realistic temperature distribution on the hot plate (with emissivity close to 1.0) indicating that it 
does not exceed much the temperature of the crucible at the emissivity sticker. The metallic surfaces of tin and aluminum do not 
allow one to detect the temperature in the infrared image because of the low surface emissivity and reflections. Such 
measurements would work better, e.g., with a crucible made of graphite and an oxide as a model material.

The discrepancy between the setpoint and actual temperature in Fig.~\ref{fig:meas} allows one to understand the plot of power 
control. The power percentage is evaluated from the on-time divided by the 20~s period, so that the 16\% power level during the 
growth phase corresponds to 3.2~s on-time and 16.8~s off-time and an average heating power of 240~W (16\% of 1500~W). 
Note that the minimum on-time was set to 2~s, so that 10\% corresponds to zero heating (e.g., after 155~min in 
Fig.~\ref{fig:meas}, when the actual cooling is slower than the specified ramp). The power data extracted from the current 
sensor shows slightly higher values than applied in the power control in Fig.~\ref{fig:meas}. This is related to the time constant 
of the active rectifier and peak detection for the current signal consisting of packets of 50~Hz sinusoidal waves.

Finally, data from the crystal weight sensor is given in Fig.~\ref{fig:meas}. All relevant phases of the growth process can be 
identified: dipping in the seed at 63~min, where the weight shortly becomes negative; growth start at 70~min, where the weight 
sharply increases; non-linear increase during the cone growth up to about 90~min; approximately linear increase in the  
remaining part, where the crystal grows with an approximately constant diameter. A weight of 255~g is reached at the end of 
the process, while the dismounted crystal showed only 216~g on a laboratory scale. When the crystal was re-attached to the 
seed mount, a weight of 241~g was measured. This discrepancy of 12\% can be explained by the previous calibration of the 
weight sensor with calibrated weights put on the top of the cell (see Fig.~\ref{fig:setup}). A later test with hanging weights in 
the range 100\ldots 500~g confirmed that the present geometry of the weight cell, metallic thread, and guiding pulleys lead to an 
apparent weight increase by 12\%. The further increase by 6\% at the end of the growth process could be related to the 
friction at the metallic thread. It should be noted that a weight cell with an appropriate operating temperature specification should 
be used, otherwise deviations as high as 1~g/\tdeg C with changing weight cell temperature above the hot plate may occur.


\section{Analytical estimations}\label{sec:analyt}

The goal of this section is to develop a basic physical picture of the growth experiment described above. The focus lays on 
\uline{macroscopic} physical phenomena such as heat transfer, electromagnetism, fluid flows, and solid stresses. The following 
paragraphs present analytical  expressions for these phenomena and discuss the relevance of various effects in the growth 
process. Although a specific CZ growth setup is discussed, most of the estimations can be easily applied to different growth 
methods and are summarized in Appendix~\ref{app:analyt}. The developed simple physical pictures will be compared to results 
from numerical models in the next section. Material properties of tin are given in Tab.~\ref{tab:mat} and are compared (for 
latter reference) to elemental semiconductors silicon (Si) and germanium (Ge).

\begin{table}[!htb]\centering
\begin{tabular}{|p{0.15\linewidth}|l|l|l|l|l|l|l|l|} \hline
\textbf{Property} & & \textbf{Units} & \textbf{Si(S)} & \textbf{Ge(S)} & \textbf{Sn(S)} & \textbf{Si(L)} & \textbf{Ge(L)} & \textbf{Sn(L)}\\ \hline
Density & $\rho$ & kg/m\tsup{3} & 2330 & 5370 & 7280 & 2560 & 5534 & 6980 \\ \hline
Heat capacity & $c$ & J/kgK & 1040 & 418 & 257 & 1004 & 358 & 257 \\ \hline
Thermal conductivity & $\lambda$ & W/mK & 23 & 17 & 60 & 62 & 48 & 32 \\ \hline
Melting point & $T_0$ & \tdeg C & 1412 & 937 & 232 & -- & -- & -- \\ \hline
Latent heat & $q_0$ & J/kg & 1.8\tdot 10\tsup{6} & 4.6\tdot 10\tsup{5} & 6.1\tdot 10\tsup{4} & -- & -- & -- \\ \hline
Emissivity & $\epsilon$ & -- & 0.46 & 0.55 & 0.07 & 0.2 & 0.2 & 0.1 \\ \hline
Thermal expansion coef. & $\beta$ & 1/K & 4.5\tdot 10\tsup{-6} & 7.5\tdot 10\tsup{-6} & 2.2\tdot 10\tsup{-5} & 1\tdot 10\tsup{-4} & 1\tdot 10\tsup{-4} & 1\tdot 10\tsup{-4} \\ \hline
Viscosity & $\eta$ & Pa\tdot s & -- & -- & -- & 5.7\tdot 10\tsup{-4} & 6.8\tdot 10\tsup{-4} & 0.002 \\ \hline
Surface tension & $\gamma$ & N/m & -- & -- & -- & 0.83 & 0.6 & 0.56 \\ \hline
Young’s modulus & $E$ & GPa & 144 & 118 & 50 & -- & -- & -- \\ \hline
Shear modulus & $G$ & GPa & 59 & 50 & 18 & -- & -- & -- \\ \hline
Electrical conductivity & $\sigma$ & S/m & 3.2\tdot 10\tsup{4} & 1\tdot 10\tsup{5} & 4\tdot 10\tsup{6} & 1.4\tdot 10\tsup{6} & 1.4\tdot 10\tsup{6} & 1.2\tdot 10\tsup{6} \\ \hline
\end{tabular}
\caption{Physical properties of Ge and Si compared to Sn, in liquid (L) and solid (S) (close to melting point) states.}
\label{tab:mat}
\end{table}

The first model addresses the \uline{heat balance in the crystal} (see, e.g. \autocite{baehr_heat_2011}) during the growth 
considering a crystal height of $H=100\textrm{~mm}$ and a diameter of $D=10\textrm{~mm}$. Heat flows into the crystal at 
the crystallization interface (area $S_0=\pi D^2/4$) due to the latent heat $Q_{lat}=S_0\rho_S q_0 v$ and temperature gradient 
in the melt $Q_{melt}=S_0\lambda_L \frac{T_m-T_0}{h}$. Assuming a melt temperature of $T_m=237\textrm{~\tdeg C}$ (see 
Fig.~\ref{fig:meas}) and height $h=10\textrm{~mm}$ we obtain $Q_{cond}=1.3\textrm{~W}$. The latent heat results in 
$Q_{lat}=2.3\textrm{~W}$ for a growth velocity of $v=4\textrm{~mm/min}$. Heat flows out of the crystal at the side surface 
(area $S_S=\pi DH$) due to radiation $Q_{rad}=S_S\epsilon_S \sigma_{sb}(T_S^4-T_a^4)$  and gas convection 
$Q_{conv}=S_S\alpha(T_s-T_a)$, where $T_a=23\textrm{~\tdeg C}$ is the ambient temperature. The crystal surface 
temperature is obviously somewhere in the range $T_0=232\textrm{~\tdeg C}$\ldots$T_a=23\textrm{~\tdeg C}$, so that we 
obtain $Q_{rad}\leq\textrm{0.7~W}$. Due to heat balance $Q_{melt}+Q_{lat}= \textrm{3.6~W} = Q_{conv}+Q_{rad}$ we can 
further estimate $Q_{conv}=\textrm{2.9\ldots 3.6~W}$ and $\alpha=\textrm{4.6~W/m\tsup{2}K}$ for $T_S=T_0$. 
Consequently, heat loss by air convection seems to dominate. Note that heat loss through conduction in the seed was neglected 
due to the small seed diameter and punctual contacts to the mount.

Electromagnetic phenomena in crystal growth are often related to the \uline{induction of heat and forces in the melt} (see, e.g. 
\autocite{lupi_induction_2015, dadzis_modeling_2012}). In the present setup, heat is generated in a resistance heater within the 
hot plate, which is supplied with 50~Hz AC current. The current $I$, voltage $U$, and power $P$ are related by 
$P=I^2_\mathrm{eff}R$ and $U_\mathrm{eff}=I_\mathrm{eff}R$. From $P=1500\textrm{~W}$ and 
$U_\mathrm{eff}=230\textrm{~V}$ we obtain $I_\mathrm{eff}=6.6\textrm{~A}$ and $R=35\,\Omega$. Measurements with 
the current sensor showed a current amplitude of 9~A, corresponding to an effective value of 6.4~A. As a theoretical exercise let 
us consider possible induction effects from the current with frequency $f=50\textrm{~Hz}$. The skin depth in liquid tin can be 
estimated as $\delta_{em}=1/\sqrt{\mu_0\pi f\sigma}= 65\textrm{~mm}$. The magnetic field is estimated assuming a circular 
current with diameter $D=120\textrm{mm}$: $B=\mu_0 I/D = 0.1\textrm{~mT}$. A first estimation (assuming expressions 
for a distinct skin effect) of the induced heat $Q_{em}=B_0^2 / 2\mu_0^2 \sigma\delta_{em} = 0.04~\textrm{W/m\tsup{2}}$ 
(or 0.6~Wm\tsup{3} if divided by skin depth) and induced force 
$F_{em}=B_0^2 / 2\mu_0\delta_{em} = 0.06\textrm{~N/m\tsup{3}}$ present practically negligible values.

Due to the temperature gradients, \uline{buoyant convection} (see, e.g. \autocite{baehr_heat_2011}) takes place both in the 
melt and in the surrounding air. The buoyancy forces $F_{buoy}=\rho g\beta\Delta T$ can be estimated to 34~N/m\tsup{3} and 
3~N/m\tsup{3} assuming temperature differences of 5~\tdeg C and 100~\tdeg C in the melt and air 
($\rho=\textrm{1~kg/m\tsup{3}}$, $\beta=\textrm{0.003~1/K}$, $\eta=2\cdot 10^{-5}\textrm{~Pa s}$, 
$c=1000~\textrm{J/kg K}$, $\lambda=0.03\textrm{~W/mK}$), respectively. To obtain a characteristic 
flow velocity, it is assumed that this force accelerates a fluid volume for a length $L$ as $u=\sqrt{2LF/\rho}$.This leads to 
0.01~m/s for the melt ($L=10\textrm{~mm}$) and 0.8~m/s for air ($L=100\textrm{~mm}$).

The \uline{free melt surface and the meniscus at the crystal} (see, e.g. \autocite{duffar_crystal_2010}) have a central role in the 
growth process. Meniscus height can be approximated with $h_m=a\sqrt{1-sin\beta_0}$, where 
$a=\sqrt{2\gamma /\rho g}=4\textrm{~mm}$ and $\beta_0$ is the angle between the vertical direction and the meniscus. 
Obviously, if the solid/liquid contact angle is approximately zero, crystal diameter increases with $\beta_0>0$, decreases 
with $\beta_0<0$, and remains constant with $\beta_0=0$. On the other hand, meniscus height is determined by the position of 
the triple line (i.e., outer edge of the crystallization interface). Consequently, an upward moving crystallization interface means an 
increasing meniscus height $h_m$, a decreasing meniscus angle $\beta_0$, leading to decreasing crystal diameter, and vice 
versa. Motion of the crystallization interface occurs if the thermal balance  $Q_{melt}+Q_{lat} = Q_{conv}+Q_{rad}$ is not 
fulfilled, so that thermal conditions are related to dynamic changes in crystal diameter. 

Finally, the heat flow through the crystal and the related temperature gradients cause \uline{thermal stresses} (see, e.g. 
\autocite{timoshenko_theory_1970}). A simple estimation of the stress level can be based on Hooke’s law: 
$\sigma_T=E\beta\Delta_r T$. However, the radial temperature difference in the crystal must be applied. From the previous 
thermal model we obtain $\Delta_r T=(Q_{rad}+Q_{conv})\cdot R/S_s\lambda=0.1~\textrm{~K}$, which leads to a thermal 
stress of 110~kPa.


\section{Numerical modeling}\label{sec:num}

The next step after a description and analysis of the physical phenomena involved in the growth process is a quantitative 
numerical model. In this section, the open source finite element program \textit{Elmer} \autocite{noauthor_elmer_nodate} is 
applied for a 2D multi-physical model. The level of description stays rather general, and the \uline{physical 
assumptions}, required \uline{input parameters}, and main \uline{physical results} are discussed without going into 
mathematical or numerical details (see \textit{Elmer Models manual} \autocite{raback_elmer_2023}). This perspective might 
well be the preferred one for practical crystal growers. Here, it allows one to focus on the topic of \uline{validation}, i.e., to 
separate this topic from verification \autocite{dadzis_validation_2017} and other more technical aspects.

The geometry and a triangular mesh are created by \textit{Gmsh} \autocite{noauthor_gmsh_nodate} using a \textit{GEO} script. 
The geometry includes tin melt and crystal, crucible, hot plate, coil below the plate (for the electromagnetic calculation) and the 
air domain with 60~cm diameter and 50~cm height. The model in \textit{Elmer} is defined using a \textit{SIF} file. The used 
\textit{Elmer} modules, applied physical assumptions and the corresponding source terms and boundary conditions are 
summarized in Tab.~\ref{tab:bcs}. Together with geometric simplifications and material properties (see Tab.~\ref{tab:mat} for 
Sn(L) and Sn(S)) these are the \textit{input} data for the model. A discussion follows in the next section, but note that similar 
approaches as shown in Tab.~\ref{tab:bcs} have been often used in the literature for crystal growth simulation.

\begin{table}[!htbp]\centering\setdefaultleftmargin{1em}{}{}{}{}{}
\begin{tabular}{|p{0.22\linewidth}|p{0.3\linewidth}|p{0.4\linewidth}|}
    \hline
\textbf{\textit{Elmer} module and main equations}  & \textbf{Physical assumptions}  & \textbf{Source terms \& boundary conditions} \\ 
    \hline
\textit{HeatSolve}: Heat conduction eq. &
\begin{minipage}[t]{\linewidth}\begin{compactitem}
\item Fixed crystallization interface (may be non-isothermal)
\item Radiation to ambient (not surface to surface)
\end{compactitem}\end{minipage}%
&
\begin{minipage}[t]{\linewidth}\begin{compactitem}
\item Homogeneous heat source in hot plate (adjusted to reach melting temperature at triple point)
\item Prescribed latent heat on crystallization interface
\item Radiation to ambient and convective cooling on solid surfaces
\end{compactitem}\end{minipage}%
    \\ \hline
\textit{MagnetoDynamics-2D}: Vector potential eq. &
\begin{minipage}[t]{\linewidth}\begin{compactitem}
\item Ring current
\item Limited air domain
\end{compactitem}\end{minipage}%
&
\begin{minipage}[t]{\linewidth}\begin{compactitem}
\item Prescribed current density in coil
\item Zero magnetic potential on outer boundary
\end{compactitem}\end{minipage}%
    \\ \hline
\textit{FlowSolve}: Navier-Stokes eq. &
\begin{minipage}[t]{\linewidth}\begin{compactitem}
\item Laminar melt flow
\item Incompressible, laminar gas flow
\item Limited gas domain
\end{compactitem}\end{minipage}%
&
\begin{minipage}[t]{\linewidth}\begin{compactitem}
\item Buoyancy forces in Boussinesq approximation
\item No-slip condition on ALL surfaces
\item Increased viscosity: 10x in melt, 100x in air
\end{compactitem}\end{minipage}%
    \\ \hline
\textit{StressSolve}: Navier eqns. &
\begin{minipage}[t]{\linewidth}\begin{compactitem}
\item Rigid seed mount
\item Isotropic material
\item Non-isothermal crystallization interface
\end{compactitem}\end{minipage}%
&
\begin{minipage}[t]{\linewidth}\begin{compactitem}
\item Thermal stress in the crystal
\item Zero displacement on seed/crystal boundary
\end{compactitem}\end{minipage}%
     \\ \hline
\end{tabular}
\caption{Model settings for a numerical 2D simulation using \textit{Elmer}.}
\label{tab:bcs}
\end{table}

\begin{figure}[!htbp]
\centering
\includegraphics[width=0.32\linewidth]{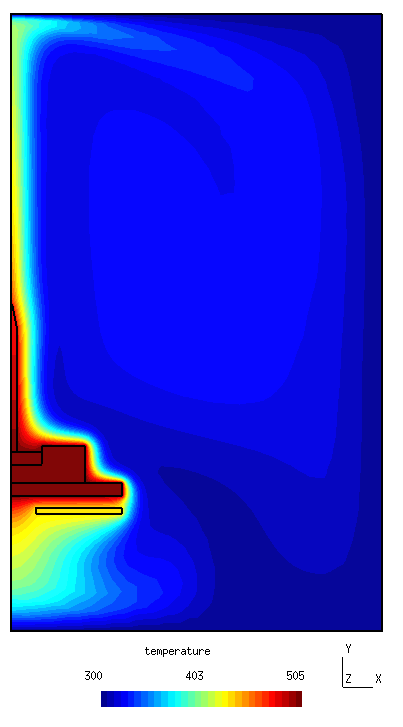}
\includegraphics[width=0.4\linewidth]{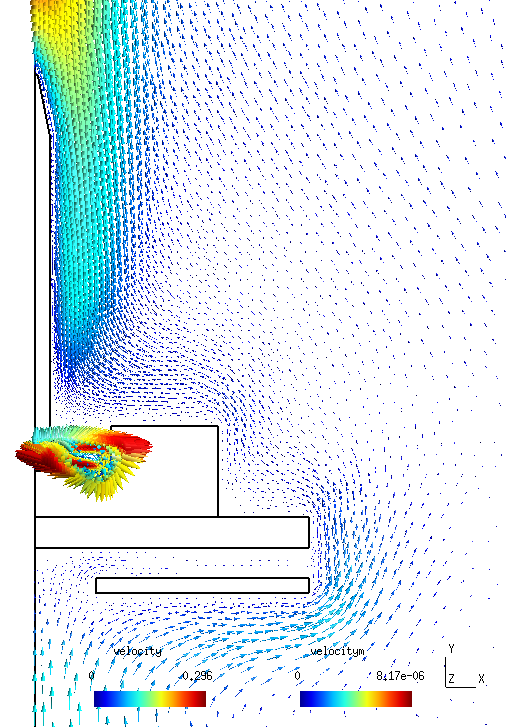}
\includegraphics[width=0.24\linewidth]{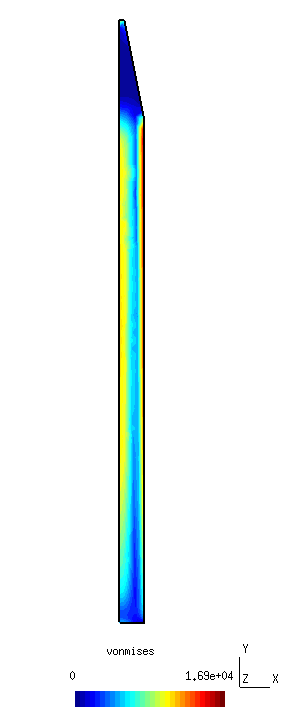}
\includegraphics[width=0.6\linewidth]{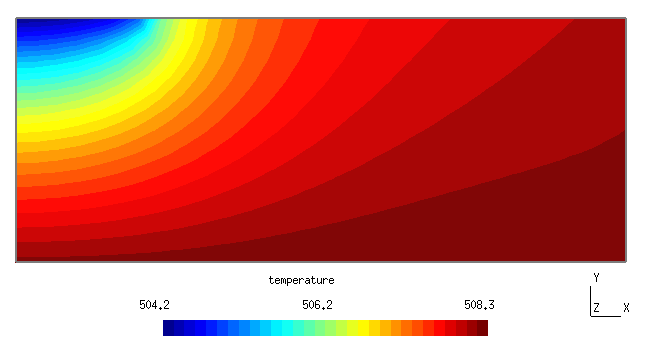}
\includegraphics[width=0.35\linewidth]{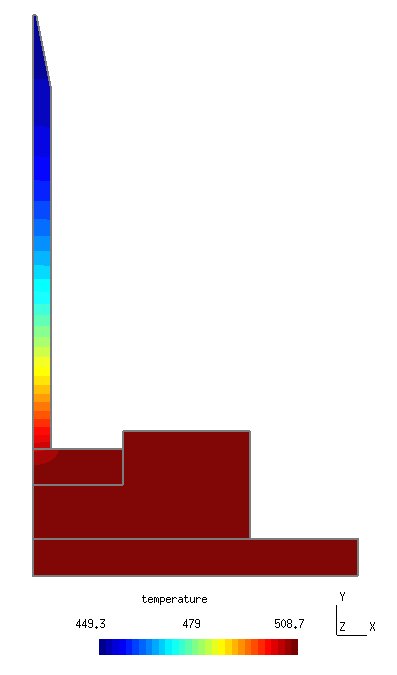}
\caption{Numerical 2D simulation with \textit{Elmer}: temperature isolines [K], velocity vectors for melt and gas flows [m/s] , 
thermal stress [Pa] in the crystal (top); thermal calculation with applied heat transfer coefficients from \textit{OpenFOAM} (bottom).}
\label{fig:elmer}
\end{figure}

Selected results from \textit{Elmer} calculations are shown in Fig.~\ref{fig:elmer}. They describe a steady-state where the 
heating power has been adjusted to obtain the melting temperature at the triple line and the latent heat corresponds to growth 
with 3~mm/min. The resulting heating power was 104~W, it decreased to 83~W without convective cooling in air and to 75~W 
with only radiation in air. These powers are smaller than 240~W measured in the experiment. Summary of heat flows on various 
surfaces in Tab.~\ref{tab:heat} indicates that heat radiation dominates heat losses from the hotplate with emissivity $
\epsilon=0.5$ and has a comparable influence to convection on other surfaces with emissivities $\epsilon\leq 0.1$. The 
calculated minimum crystal temperature was at least 209\tdeg C.

The calculation of melt and gas flows in \textit{Elmer} had convergence issues even in unsteady calculations, and solutions could 
be obtained only with artificially increased viscosities (improving the numerical stability) of 10x in the melt and 100x in the air. 
Flow velocity reaches 0.01~mm/s in the melt and 0.3~m/s in the air, where the former value is even 3 orders of magnitude 
smaller than estimated in Sec.~\ref{sec:analyt}. Although the temperature differences in the aluminum crucible are below 1~K, 
the direction of the resulting gradient at the melt side obviously determines flow direction. A heat flow from the melt into the 
crucible wall leads to an upward buoyant flow at the melt side. The calculated gas flow pattern consists of an upward buoyant 
flow along the hot crystal surface as expected.

\begin{figure}[!htb]
\centering
\includegraphics[width=0.4\linewidth]{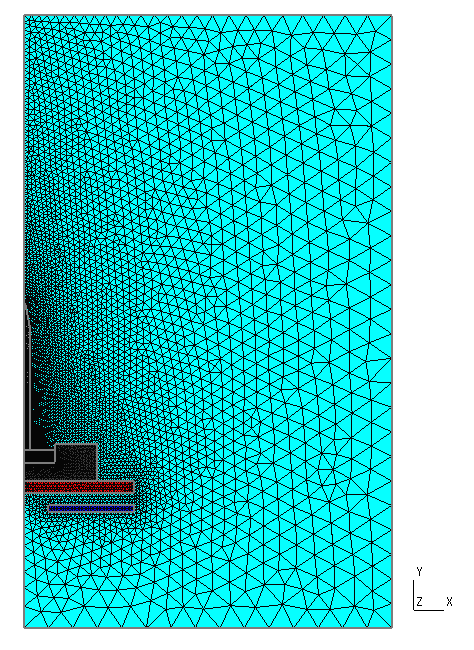}
\includegraphics[width=0.54\linewidth]{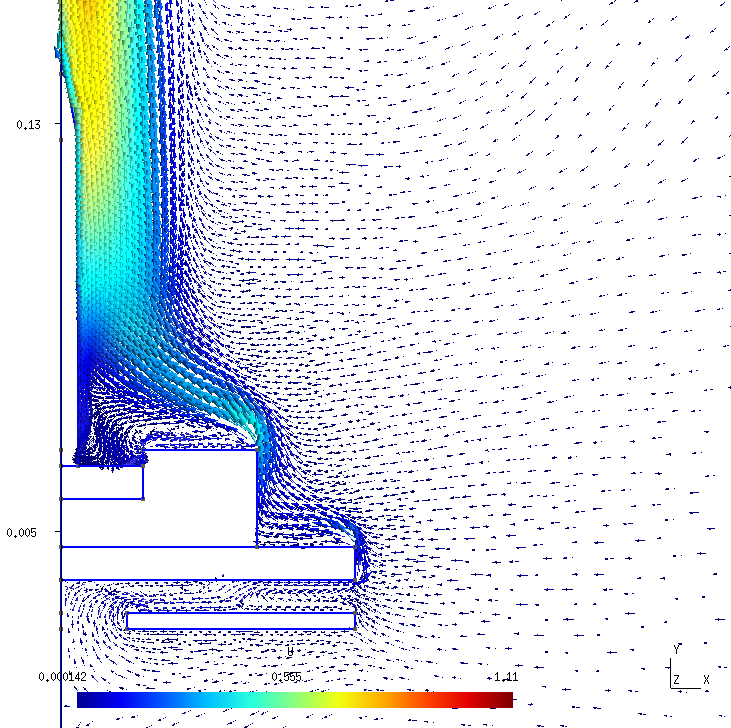}\\[1em]
\includegraphics[width=0.48\linewidth]{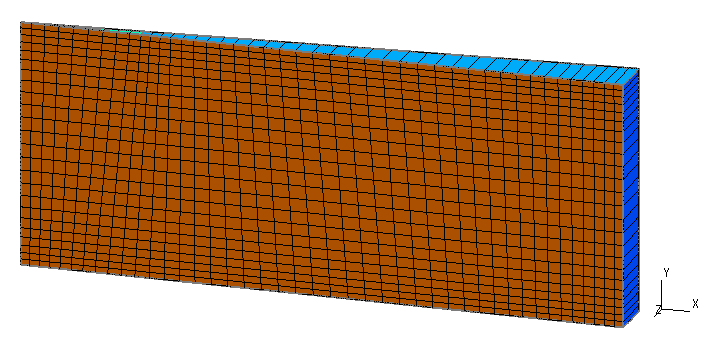}
\includegraphics[width=0.48\linewidth]{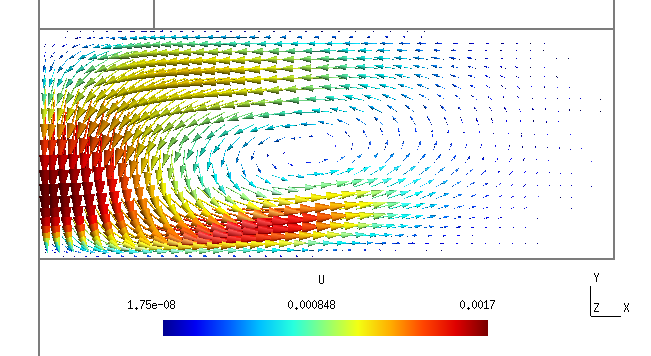}
\caption{Numerical 2D simulation with \textit{OpenFOAM}: grid (left), velocity vectors [m/s] for melt and gas flows (right).}
\label{fig:of}
\end{figure}

Comparative 2D calculation for the melt and gas flows were performed using \textit{OpenFOAM} 
\autocite{noauthor_openfoam_nodate}. The triangular mesh of the gas domain was rotated (extruded) in \textit{Gmsh} to a 
5\tdeg ~wedge creating prismatic 3D elements, see Fig.~\ref{fig:of}. Simplified boundary conditions with constant surface 
temperatures (gas domain: 494~K at crystal, 505~K at other solids; melt domain: 505~K at crystal, 508~K at crucible bottom, zero 
heat flux at other boundaries) from \textit{Elmer} calculation were applied. Unsteady calculations using the incompressible 
\textit{buoyantBoussinesqPimpleFoam} solver with upwind schemes for the convective terms typically reached a stationary state 
before 100~s. The global gas flow pattern and the characteristic velocity shown in Fig.~\ref{fig:of} remained similar to 
Fig.~\ref{fig:elmer}, although some recirculation vortices appeared at the melt free surface. However, the convective heat fluxes were 
much higher than in the previous \textit{Elmer} calculation. The heat transfer coefficients (HTC) were estimated as $
\alpha=3\textrm{~W/m\tsup{2}K}$ on melt free surface and $\alpha=6\textrm{~W/m\tsup{2}K}$ on other solid surfaces. A thermal 
calculation with \textit{Elmer} using these coefficients led to much higher convective heat losses (see Tab.~\ref{tab:heat} and 
Fig.~\ref{fig:elmer} (bottom) for resulting temperature fields) and a total heating power of 173~W.  Obviously, the artificially 
increased viscosity may cause underestimated convective heat exchange. The melt flow calculation in \textit{OpenFOAM} reaches a 
velocity of 2~mm/s, which is much higher than in \textit{Elmer}.

\begin{table}[!htb]\centering
\begin{tabular}{|l|l|l|l|l|} \hline
\textbf{Surface} & \textbf{With conv.} & \textbf{No conv.} & \textbf{Rad. only} & \textbf{With HTC} \\ \hline
Crystal ($\epsilon=$0.07) & 1.7 & 1.4 & 0.8 (0.8) & 4.3 \\ \hline
Melt ($\epsilon=$0.1) & 0.7 & 0.6 & 0.5 (0.6) & 2.2 \\ \hline
Crucible ($\epsilon=$0.05) & 13.2 & 6.6 & 5.0 (3.3) & 32.5 \\ \hline
Hotplate ($\epsilon=$0.5) & 89.5 & 76.1 & 70.2 (72) & 136.4 \\ \hline
\multicolumn{1}{|r|}{\textbf{Sum}} & 105.1 & 84.7 & 76.5 & 175.4 \\ \hline
\multicolumn{1}{|r|}{\textbf{Heating}} & 103 & 83 & 75 & 173 \\ \hline
Crystal min. T & 482~K & 485~K & 496~K & 449~K\\ \hline
\end{tabular}
\caption{Heat flows in [W] in various calculations with \textit{Elmer} evaluated using \textit{SaveScalars} procedure and 
\textit{Diffusive flux} operator with 2nd order elements. In the case ``Rad. only'' analytical estimation is given in brackets ().}
\label{tab:heat}
\end{table}

The electromagnetic calculation considers an AC current below the hotplate producing a magnetic field, which generates eddy 
currents in electrically conducting bodies. If a non-conducting crucible material is considered, the liquid tin melt is subject to a 
magnetic flux density of 0.2~mT, induced heat power up to 0.03~W/m\tsup{3}, and induced Lorentz force densities up to 
0.0001~N/m\tsup{3}. Obviously the inductive effects are negligible both for the thermal and melt flow calculation. \uline{As a 
theoretical exercise}, another case was calculated for an inductive hotplate with a frequency of 22~kHz and a current of 30~A in 
20 windings. A graphite crucible with an electrical conductivity of 5\tdot 10\tsup{4}~S/m leads to a magnetic flux density of 
11~mT, induced heat power up to 2\tdot 10\tsup{6}~W/m\tsup{3} in the crucible and 8\tdot 10\tsup{5}~W/m\tsup{3} in the 
melt, induced Lorentz force densities in the melt up to 10\tsup{3}~N/m\tsup{3}. In this case, significant electromagnetic effects 
can be expected.

Thermal stress in the crystal was calculated assuming isotropic elastic properties. As shown in Fig.~\ref{fig:elmer}, the von 
Mises stress first decreases radially, but then increases and reaches a maximum of 2\tdot 10\tsup{4}~Pa at the crystal rim. The 
stress components $\sigma_{ij}$ have radially monotonous distributions with values up to 10\tsup{4}~Pa. While the case in 
Fig.~\ref{fig:elmer} has a rather low stress level at the crystallization interface, artificial local maxima may occur due to local 
thermal gradients caused by the prescribed flat interface shape.

Note that all simulations described here were done on the Raspberry Pi computer used for process monitoring in 
Sec.~\ref{sec:exp}. This approach would allow the implementation of advanced algorithms for process control involving both in-situ 
measurement data and live numerical models. The source code for \textit{Elmer} and \textit{OpenFOAM} models  is 
available as open source (see App.~\ref{app:online}).


\section{Discussion}\label{sec:disc}

The analytical models in Sec.~\ref{sec:analyt} and numerical models in Sec.~\ref{sec:num} both describe a CZ growth process 
in terms of selected \textit{physical pictures}. In the following sections, main results from both approaches are discussed and 
compared to achieve the following goals:\vspace{-\parskip}
\begin{compactitem}
\item Develop a physical understanding of the CZ growth process on a macroscopic level.
\item Identify open questions in the modeling methods and propose solutions in terms of verification and validation.
\end{compactitem}

The analysis of \uline{heat balance in the crystal} in Sec.~\ref{sec:analyt} arrived at the conclusion that heat loss due to air 
convection at the crystal surface exceeds heat radiation several times. While first numerical results in \textit{Elmer} 
indicated an opposite relation, a closer analysis of air convection confirmed the dominating influence of convection even if crystal 
surface temperature drops only about 20~K below the melting point. It can be concluded:\vspace{-\parskip}
\begin{compactitem}[\tarr]
\item The effect of convective cooling may be grossly underestimated in simplified models of gas flows with artificially increased 
viscosities, under-resolved boundary layers, etc. Verification using benchmark cases and validation using measurements of 
crystal/seed temperatures is recommended.
\item Accuracy of integral heat fluxes in numerical models should be verified with analytical solutions. Use of second order elements in \textit{Elmer} were crucial to obtain accurate results.
\end{compactitem}

In the sense of a \uline{global heat balance} of the growth setup, the sum of all heat losses should be equal to the heating power 
(plus a small contribution by the latent heat). The numerically calculated power of 173~W is smaller than the measured power of 
240~W by almost 30\%. To explain such deviations:\vspace{-\parskip}
\begin{compactitem}[\tarr]
\item Heat losses missing in the numerical model could be identified (i.e., validated) in the experimental setup using thermal 
imaging and additional temperature or heat flux measurements to find theoretically unexpected \textit{cold spots}.
\item Material properties with a large degree of uncertainty and large influence on the heat balance (e.g., emissivity of solid 
surfaces) should be validated in dedicated experiments, e.g., using pyrometric measurements.
\end{compactitem}

The measurement of the AC heater current allowed us to calculate the \uline{heat and force induction in the melt}. While the 
analytical and numerical values of the magnetic field agree well, the induced heat density and force density are smaller by 1–3 
orders of magnitude in the numerical simulation for the 50~Hz case. While the induction effects remain negligible both in 
simulation and estimation, for the application of higher frequencies such as 22~kHz, a closer analysis of such discrepancies is 
recommended.\vspace{-\parskip}
\begin{compactitem}[\tarr]
\item In addition to the heater power, the measurement of the heater current or the related magnetic field is a prerequisite for 
further analysis of induction effects. This can be seen as validation of heater models for complex heater shapes and temperature-
dependent properties.
\item The numerical calculation of heat and force induction should be verified using analytical solutions for simple geometries.
\end{compactitem}

\uline{Buoyant convection} in the air is obviously responsible for the strong effect of convective cooling discussed above. Despite 
the strongly simplified boundary conditions for the air flow in the numerical models with \textit{Elmer} and \textit{OpenFOAM}, 
the calculated typical velocities agree with the analytical estimation. On the contrary, the numerically calculated velocity for the 
melt flow is nearly 3 orders of magnitude smaller in \textit{Elmer}. The following general aspects may be emphasized here:\vspace{-\parskip}
\begin{compactitem}[\tarr]
\item Artificial increase of fluid viscosity (as tested in \textit{Elmer}) to improve the numerical convergence may lead to large deviations in the calculated velocity magnitude and heat transfer coefficients.
\item Boundary conditions for gas flows should be validated with velocity or pressure measurements to enable physically 
meaningful solutions also for strongly simplified models (such as a closed box with cold walls).
\item High thermal conductivity of the melt may lead to very low thermal gradients, so that even small changes in local thermal 
conditions may influence the resulting buoyant flow pattern. Both a validation of the thermal model with accurate temperature 
measurements and velocity measurements in the melt are recommended.
\end{compactitem}

In the numerical simulation with \textit{Elmer}, a flat free melt surface with a fixed triple line position was assumed. 
Consequently, the \uline{liquid meniscus at the crystal} was neglected. Analytical estimation leads to a meniscus height of 4~mm 
for a meniscus angle of 0\tdeg . The understanding of the dynamic changes of crystal diameter and development of related 
analytical or numerical models would be hardly possible without considering the following aspects:\vspace{-\parskip}
\begin{compactitem}[\tarr]
\item A validation using optical observation of the triple line region is crucial. The exact relationship between growth angle and 
meniscus height may be not only material-dependent but also process dependent. 
\item A fixed triple line and neglected undercooling in the simulation may lead to unphysical results such as artificial thermal 
gradients. Such assumptions should be validated with measurements of temperature and crystallization interface shape. 
\end{compactitem}

\uline{Thermal stress} in the crystal is a direct consequence of the temperature non-linearity and the mechanical constraints. The 
analytical estimation arrived at about 100~kPa and the numerical calculation at 10~kPa, the latter with a relatively constant radial 
profile over a large part of the crystal height. This can be probably explained with the trivial analytical formula. Nevertheless, 
some assumptions in numerical models may need a closer attention:\vspace{-\parskip}
\begin{compactitem}[\tarr]
\item The thermal model of the crystallization interface (e.g., deviations from an isotherm) may cause artificial local stresses. 
This model should be verified with analytical or accurate numerical solutions.
\item The non-linearity of the temperature distribution may be enhanced by local thermal asymmetries, small deviations from the 
idealized crystal shape and other hardly measurable factors. Here a validation using a dummy crystal allowing for in-situ stress 
analysis in a comparable thermal process could be helpful. 
\end{compactitem}

Although the discussed aspects above mostly arose from the analysis of a very specific CZ growth process, a wider literature 
study about modeling of CZ growth demonstrated similar open questions in model validation 
\autocite{enders-seidlitz_development_2022}. Furthermore, one might expect similar issues in many other growth processes 
(and comparable multi-physical processes) sharing the same basic physical phenomena. This conclusion has served as a 
motivation for the NEMOCRYS project \autocite{noauthor_next_nodate}, where the following strategy is being applied:\vspace{-\parskip}
\begin{compactenum}
\item Identify open questions in models of crystal growth processes.
\item Develop model experiments with extensive in-situ measurements.
\item Build new validated coupled models for crystal growth.
\end{compactenum}
This study presented a simple model experiment for CZ growth, basic techniques for in-situ measurements, and simplified 
numerical models based mainly on theoretical considerations. The next iteration with respect to CZ model setups and in-situ 
measurements \autocite{enders-seidlitz_development_2022} as well as validated numerical models  
\autocite{enders-seidlitz_model_2022} has already been published. It is also planned to extend this approach to other crystal 
growth methods such as the floating zone process with inductive and/or optical heating.

Finally, it should be noted that Sn has been already used as a \textit{model material} for CZ growth in the literature 
\autocite{ekhult_czochralski_1986}. Tab.~\ref{tab:mat} compares material properties relevant for a macroscopic analysis of the 
growth process between Sn and elemental semiconductors Si and Ge, where the CZ growth has a great practical significance. It 
can be seen that some properties agree relatively well, while others such as the latent heat differ more than by an order of 
magnitude. A more exact analysis of the physical similarity between the present experiment with Sn and CZ growth of Si or Ge on 
small or large scales would be possible using analytical estimates (as in Sec.~\ref{sec:analyt}) and dimensionless numbers \autocite{dadzis_modeling_2012, dadzis_model_2020}.


\section{Conclusions and outlook}\label{sec:conc}

A new low-cost experimental setup with open hardware and open source software has been developed for the CZ crystal growth 
process of model materials up to about 350~\tdeg C. The compact size of the setup, its automation capabilities as well as detailed 
in-situ measurements enable the application of this setup not only for demonstration and educational purposes but also for 
small-scale conceptual studies.

In the present study, tin crystals with diameters of about 12~mm and lengths up to 320~mm were grown. In contrast to the 
historic experiments by Jan Czochralski in 1916, the main focus was not on the determination of the highest achievable growth 
rates but rather on a reproducible growth process with comprehensive in-situ observation to extract the typical physical 
phenomena of CZ growth in the sense of a model experiment. 

The analysis of the experiment allowed to set up analytical and numerical models describing the physical aspects of heat transfer, 
electromagnetism, melt and gas flows as well as solid stresses in CZ growth. The following discussion of the theoretical and 
experimental data demonstrated a new strategy for the development and validation of macroscopic crystal growth models. 

Following the proposed strategy, new model experiments will be developed for various crystal growth processes in the 
NEMOCRYS project. They will be applied to address the many open questions in crystal growth simulation and in the complex 
interaction of various physical phenomena. In summary, Jan Czochralski’s experiment in its simplicity teaches us to separate the 
core scientific principle of the CZ method from the following technological advances. It is helpful to look for deep answers to 
simple (modeling) questions before trying simplified answers to deep (modeling) questions.


\section*{Acknowledgments}

Arved Wintzer (IKZ) continuously supported the analysis of analytical and numerical results in this work. Peter Gille (Ludwig 
Maximilian University of Munich), Olf Pätzold (Technical University Bergakademie Freiberg) and Christo Guguschev (IKZ) are 
gratefully acknowledged for many useful discussions about model materials and model experiments. Frank Kießling (IKZ) provided 
helpful comments on the manuscript. This study received funding from the European Research Council (ERC) under the European 
Union’s Horizon 2020 research and innovation programme (grant agreement No 851768).


\printbibliography


\appendix

\section{Online materials (in preparation)}\label{app:online}


Schematics and tests of the Arduino-based control:
\url{https://github.com/nemocrys/czdemo-exp}

Source code for the Arduino-based control:
\url{https://github.com/nemocrys/czdemo-exp/tree/master/arduino}

Python scripts for visualization of camera images and measurement data as well as for preparing and uploading growth recipes on 
the Raspberry Pi:
\url{https://github.com/nemocrys/czdemo-exp/tree/master/raspberrypi}

Interactive simulation of the Czochralski growth proces in Python: 
\url{https://github.com/nemocrys/crystal-game}

Source code for the solvers in \textit{Elmer} and \textit{OpenFOAM}:
\url{https://github.com/nemocrys/czdemo-sim}

\section{Analytical estimation of crystal growth processes}\label{app:analyt}

\includegraphics[width=1.0\linewidth]{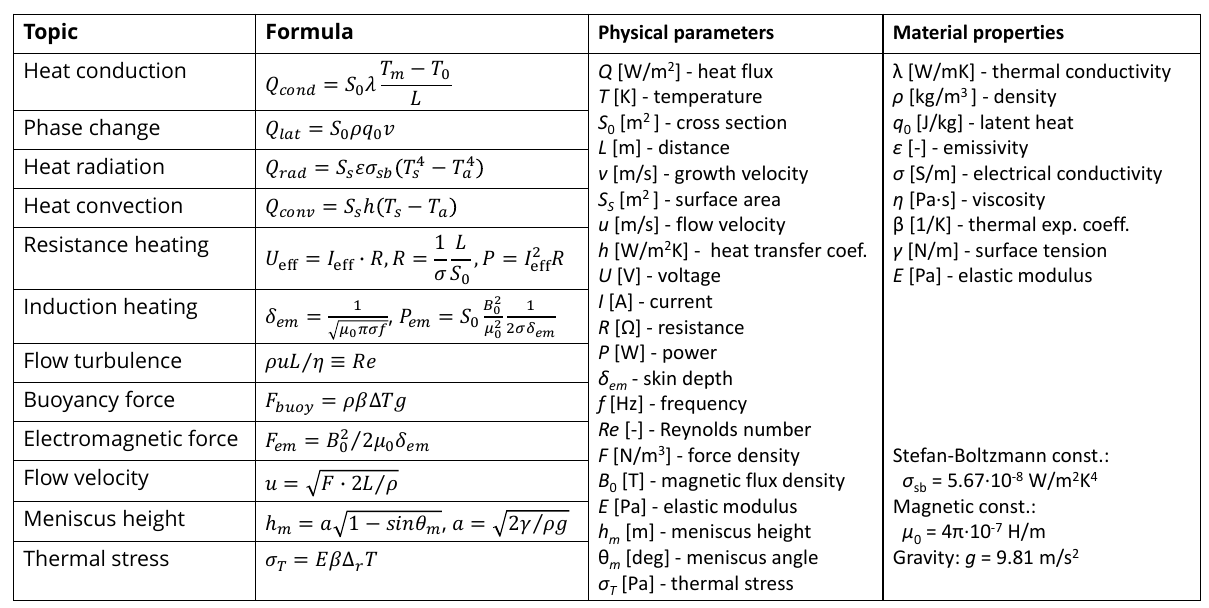}

\end{document}